\documentclass[aps,pra,amsmath,amssymb,onecolumn, notitlepage]{revtex4-1}
\usepackage{amsmath,epsfig,amssymb}
\usepackage{bbm}
\usepackage{times}
\usepackage{color}
\usepackage{amsthm}
\usepackage{color}
\usepackage[justification=centering]{subfig}

\newcommand{\todo}[1]{}
\renewcommand{\todo}[1]{{\color{red} TODO: {#1}}}
\renewcommand\Re{\operatorname{Re}}
\renewcommand\Im{\operatorname{Im}}

\DeclareMathOperator{\Tr}{Tr}

\newtheorem{theorem}{Theorem}[section]

\begin{document}

\title{Orientation Statistics and Quantum
Information\footnote{DISTRIBUTION
STATEMENT A - APPROVED FOR PUBLIC RELEASE; DISTRIBUTION IS UNLIMITED.}}
\author{Kevin Schultz}
\affiliation{
 Johns Hopkins University Applied Physics Laboratory\\
 11100 Johns Hopkins Road, Laurel, MD, 20723, USA
}

\begin{abstract}
    
    Motivated by the engineering applications of uncertainty quantification, in
    this work we draw connections between the notions of random quantum states
    and operations in quantum information with probability distributions
    commonly encountered in the field of orientation statistics.  This approach
    identifies natural probability distributions that can be used in the
    analysis, simulation, and inference of quantum information systems.  The
    theory of exponential families on Stiefel manifolds provides the
    appropriate generalization to the classical case, and fortunately there are
    many existing techniques for inference and sampling that exist for these
    distributions.  Furthermore, this viewpoint motivates a number of
    additional questions into the convex geometry of quantum operations
    relative to both the differential geometry of Stiefel manifolds as well as
    the information geometry of exponential families defined upon them.  In
    particular, we draw on results from convex geometry to characterize which
    quantum operations can be represented as the average of a random quantum
    operation.

\end{abstract}

\maketitle

\section{Introduction}

Consider some arbitrary classical (i.e., non-quantum) experimental procedure,
wherein experiments are performed and measurements are taken to quantify some
characteristic of some component.  Suppose this component is to be included as
sub-component of a larger system and an engineer is charged with performing an
analysis (likely via computer simulation) of how variability of the sub-components
affects the overall system performance.  Such studies are commonly denoted
sensitivity analysis, risk analysis, or uncertainty quantification
\cite{iman1988investigation,saltelli2000sensitivity} and such techniques are of
practical concern in a variety of contexts.
Suppose the engineer tracks down the experimental results for a given component
(e.g., length tolerances for some packaging) and is able to obtain the sample
data generated by the experimentalist, or more likely, the reported mean $\mu$
and variance $\sigma^2$.  In either case, it is likely that the engineer will
model the distribution of the characteristic using the normal distribution
$\mathcal{N}(\mu,\sigma^2)$, and indeed, there is mathematical reasoning to do
so \cite{jaynes1957information}, beyond any computational reasons for doing so.
The engineer might use samples from this distribution as part of a Monte Carlo
analysis, or perhaps the functional form of the probability distribution
function can be used as part of a closed-form analysis.

A similar situation exists in quantum information, but it is often unremarked.
On one hand, the evolution of an open quantum system can often be treated as a
noisy evolution, through the stochastic Liouville equation
\cite{kubo1963stochastic}.  This type of description applies to some of the
most common types of qubits and their predominant decoherence mechanisms (see,
e.g. Refs. \cite{Kubo1957, Schulten1978, ernst1987principles, Schneider1998,
Grigorescu1998, abergel2003, Cheng2004, Wilhelm2007}).  On the other hand,
especially in the context of quantum gates or circuits, we speak of ``the
superoperator'' or ``the error channel'' for a given quantum operation,  and
not a stochastic process. Unlike the classical case, it is impossible to
recover the quantum operation associated with an individual quantum trajectory,
only measurements collected at the end of an experiment, which individually
contain only partial information about the average quantum channel that
generated the collections of measurements.  Thus, techniques for converting
sets of quantum measurements into estimates of quantum states or channels,
called tomography, are essentially computing estimates of the average
quantities of interest, in much the same way that descriptive statistics such
as the mean and variance are used in classical contexts. This suggests that a
rigorous statistical approach must be developed in order to develop accurate
assessments of the impacts of noise and imperfections on quantum systems, when
only sample averages of an underlying probability distribution are available.

Here, we develop such a statistical approach applicable to quantum information by
showing the relationships between a number of probability distributions common
in the field of directional or orientation statistics
\cite{mardia2009directional,chikuse2012statistics}, but somewhat obscure
outside of this field, and a number of relevant structures in quantum
information.  In particular, we relate the parameters of these
distributions to certain forms of quantum operations.  The hope is that these
distributions can be used both in closed-form and Monte Carlo analysis in the
simulation of quantum circuits, as well a providing additional
foundation for different inference problems concerning quantum systems.

In the discussion below, we first cover a number of relevant forms of quantum
operations that we will later draw connections to from the field of orientation
statistics.  Next, we introduce the concept of an exponential family, which is
a geometric concept that generalizes many of the familiar properties of normal
random variables, such as the \textit{sufficiency} of the sample mean and
variance fully describing the sampled data, as well as the maximum entropy property
that the normal distribution exhibits among all random variables with a fixed
variance.  The preliminary discussion is concluded by a brief overview of
Stiefel manifolds, which will be the sample space (instead of Euclidean space)
on which the relevant probability distributions will be defined.

\subsection{CPTP Maps}
In quantum information, a quantum state is represented by a density operator
$\rho$, where $\rho\in\mathbb{C}^{N\times N}$ is a positive semi-definite,
Hermitian matrix with $\Tr(\rho) =1$ \cite{nielsen2010quantum}.    Quantum
operations are then completely positive, trace-preserving (CPTP) maps
\cite{nielsen2010quantum}.  Here, we will make the additional assumption that
the quantum maps of interest map to density operators of the same dimension as
the input dimension, but this can be generalized.  Below we will summarize some
relevant properties and equivalent representations of CPTP maps that will help to
draw the connection to the field of orientation statistics.  More information
about various forms of CPTP maps can be found in e.g.
\cite{fujiwara1999one,bruzda2009random,wood2014tensor}.

Choi's theorem on completely positive (CP) maps \cite{choi1975completely}
states that any CP map $\Phi(\rho)$ can be written in the Kraus form
\begin{equation}
	\Phi(\rho)=\sum_{i=1}^m A_i\rho A_i^\dagger
\end{equation}
where $A_i\in\mathbb{C}^{N\times N}$ and $m$ is said to be the Kraus rank of
the map $\Phi$ and $m\leq N^2$.  The added characteristic of trace preserving (TP)
is equivalent to $\sum_iA_i^\dagger A_i \triangleq \mathbbm{1}_N$.  Note that the Kraus
form is not unique, that is, different sets of Kraus operators $A_i$ can create
equivalent maps.

Let $|\cdot\rangle\rangle$ denote the vectorization operator, and
$\langle\langle\cdot|=|\cdot\rangle\rangle^\dagger$.  From the Kraus form, two
additional representations of CPTP maps can be defined, the Choi matrix form
\begin{equation}\label{eq:choi}
	\Lambda = \sum_{i=1}^m|A_i\rangle\rangle\langle\langle A_i|\,,
\end{equation}
and the Liouvillian superoperator (or dynamical matrix form)
\begin{equation}
	\mathcal{L} = \sum_{i=1}^m A_i^*\otimes A_i\,,
\end{equation}
where $*$ denotes entry-wise conjugation, not conjugate transposition
($\dagger$).
These forms are of course equivalent and are related by a reshuffling
involution \cite{bruzda2009random,wood2014tensor}.

Let $\Tr_{2}$ denote the partial trace over the second Hilbert space in a
composite system, so that $\Tr_2(A\otimes B) =A$.  The TP property implies that
$\Tr_2(\Lambda) = \mathbbm{1}_N$ (and thus $\Tr(\Lambda)=N$) and it is obvious
from the form in (\ref{eq:choi}) that $\Lambda$ is a positive semi-definite
Hermitian matrix.  As (\ref{eq:choi}) has the same functional form as a density
operator on $\mathbb{C}^{N^2\times N^2}$, there is an isomorphism between
$\Lambda/N$ and density operators on the higher dimensional space $N^2$, called
the \textit{Jamio\l{}kowski} isomorphism \cite{jamiolkowski1972linear}.  Since
$\Lambda$ is a Hermitian positive semi-definite matrix, we can diagonalize it
as $\Lambda = KDK^\dagger$ where $K$ is a unitary matrix (in particular, its
columns are orthonormal) and $D$ is a diagonal matrix whose entries
$D_{ii}\in[0,N]$ and $\sum D_{ii} = N$.  Letting $|K_i\rangle\rangle$ denote
the $i$th column of the above decomposition, $\Lambda$ can be used to define
``canonical'' Kraus operators $\sqrt{D_{ii}}K_i$.

While $\Lambda$ enjoys a number of useful structural properties, the
Liouvillian superoperator is convenient for the propagation of state, as
$\mathcal{L}(\rho) = \mathcal{L}|\rho\rangle\rangle$.  In this work, the
primary use of the superoperator will be as it relates to the Pauli transfer
matrix or affine form \cite{fujiwara1999one}.  For a single qubit system ($N=2$),
the Pauli transfer matrix is the Liouvillian superoperator expressed in the
basis spanned by the vectorized Pauli matrices $|\sigma_I\rangle\rangle$, $|\sigma_X\rangle\rangle$,
$|\sigma_Y\rangle\rangle$, $|\sigma_Z\rangle\rangle$.  Higher dimensional analogues exist
\cite{fujiwara1999one}, but the key concept is that states $\rho$ (in say, the
computational basis) are mapped to Bloch vectors $\varphi\in\mathbb{R}^{N^2-1}$
with $||\varphi||_2\leq1$ and unitary quantum operations are represented as
rotations (elements of $SO(N^2-1)$) of $\varphi$ along a hypersphere.  The
affine form of general CPTP maps can include two rotation components, a
contraction component, and linear shift, but the focus here will be on affine
representations of unitary operations.

\subsection{Sufficient Statistics and Exponential Families}

A probability distribution $p(x;\theta)$ parameterized by $\theta$ is said to be an
exponential family if $p(x;\theta)$ can be expressed as
\begin{equation}
    p(x;\theta) = \exp(\langle\theta,T(x)\rangle - \psi(\theta) + \kappa(x))\,,
\end{equation}
where
\begin{itemize}

    \item $\theta$ are the natural parameters

    \item $T(x)$ are the sufficient statistics
        \cite{koopman1936distributions,pitman1936sufficient}, typically linearly independent

    \item $\psi(\theta)$ is the log-normalizer which causes $p(x;\theta)$ to integrate to one

    \item $k(x)$ is the carrier measure that determines the support of the
	    distribution, for example on the positive reals or some manifold
	    embedded in Euclidean space

\end{itemize}

Exponential families play a prominent role in statistics
\cite{amari2007methods,barndorff2014information}, particularly in the context
of maximum-likelihood estimation.  Many common distributions such as Gaussian,
exponential, Bernoulli, etc., are exponential families.   Two basic facts of
exponential families motivate their study here.  The first is that the
so-called expectation parameters $\eta$ of a probability distribution are
defined by $\eta = E_{p(x;\theta)}[T(x)]$ and are in one-to-one correspondence with
the natural parameters $\theta$, meaning that any estimates $\hat{\theta}$
of $\theta$ are functions of sample averages of $T(x)$
\cite{amari2007methods,barndorff2014information}.  This is the key principle
that allows for samples from a normal distribution to be completely summarized
by the mean and variance of the data, and is essentially a unique property to
exponential families.

The second motivating fact is that among all possible exponential families with
sufficient statistics $T'$ that contain $T$, the exponential family whose
sufficient statistics are just $T$ maximizes the entropy $E[-\log(p(x))]$
\cite{amari2007methods}. This property is one reason why it is reasonable to
assume normal distributions on some random parameter when only a mean and
variance are reported without any additional information, such as a Bayesian
prior or valid range on the parameter.  
Note that any two sets of statistics
that span the same set in function space define the same distribution and will
enjoy the same maximum entropy property. For example, this is why mean $E[x]$
and variance $E[(x-E[x])^2]$ or mean and second moment $E[x^2]$ will both lead
to the normal distribution (when the domain is the entire real line).

In quantum information, we often speak of \textit{the} superoperator (or other
equivalent representation) associated with a quantum channel, when in physical
reality various external sources (generally grouped into semiclassical noise or bath) result
in a (seemingly) random quantum operation even when the input control pulses
are meant to be identical.  Furthermore, given the mathematics of quantum
measurement and tomography, it is not possible to measure individual quantum
trajectories and produce a sample of superoperators to apply statistical
techniques to.  Instead, the tomographic process produces an estimate of an
average superoperator $\hat{\mathcal{L}}$ or an equivalent representation \cite{Chuang1997, Merkel2013, blume2016certifying}.  The remainder of this paper discusses probability
distributions for random quantum states and CPTP maps for whom the statistic
$\hat{\mathcal{L}}$ (or an equivalent representation) is sufficient in the
sense described above.

\subsection{Stiefel Manifolds}\label{sec:stiefel}
Suppose $X\in\mathbb{F}^{n\times k}$
($\mathbb{F}=\mathbb{R}$,$\mathbb{C}$,$\mathbb{H}$) with $X^\dagger X=\mathbbm{1}_k$,
then $X$ is called a $k$-frame of orthogonal vectors in $\mathbb{F}^n$.  The set of
all such $X$ forms the Stiefel manifold $V_k(\mathbb{F}^n)$.  In particular:
\begin{itemize}
    \item  $V_n(\mathbb{R}^n)\cong O(n)$, $V_n(\mathbb{C}^n)\cong U(n)$
    \item  $V_1(\mathbb{R}^n)=S^{n-1}$, $V_1(\mathbb{C}^n)=S^{2n-1}$ ~ ~ (i.e., unit
        spheres)
    \item  $V_{n-1}(\mathbb{R}^n)\cong SO(n)$, $V_{n-1}(\mathbb{C}^n) \cong
	    SU(n)$
\end{itemize}
Thus, for random quantum operations that are unitary, $V_{n-1}(\mathbb{C}^n)$ is
of interest.  Furthermore, $V_{n-1}(\mathbb{R}^n)$ is relevant for unitary
rotations both from the Bloch representation of the operation (i.e., the PTM),
and also from decomposing $U$ into its real and imaginary parts, constructing a
real-valued matrix.  Fortunately, Stiefel manifolds are the natural sample
space for generalizations of directional statistics (often called orientation
statistics) \cite{mardia2009directional,chikuse2012statistics} so a number of
exponential families have been defined on Stiefel manifolds, along with
inference \cite{mardia2009directional,chikuse2012statistics,sei2013properties}
and methods for generating random samples
\cite{mezzadri2006generate,hoff2009simulation,kent2013new}.

Since all Stiefel manifolds are compact, they admit a Haar measure and can thus
be sampled uniformly.  This fact will be exploited in the discussion below, and
here for completeness we describe a known method for uniform sampling of real
and complex Stiefel manifolds \cite{chikuse2012statistics}.  Let
$X\in\mathbb{R}^{m\times n}$, $m\leq n$ be matrix whose entries are independent
samples from a real Gaussian distribution with zero mean and unit variance,
then $X$ is a sample of the real-valued \textit{Ginibre} ensemble
\cite{mehta2004random}. By taking the $QR$-decomposition of $X$, the polar
matrix $Q$ is an element of $V_m(\mathbb{C}^n)$ (see
\cite{mezzadri2006generate} for some technical implementation issues).
Similarly, if $X$ and $Y$ are independent samples from the Ginibre ensemble,
then the $QR$ decomposition of $Z=X+iY$ yields uniform random elements in
$V_m(\mathbb{C}^n)$.

In addition to the spaces $U(N)$, $SU(N)$, and $SO(N^2-1)$ which naturally arise in
the context of quantum operations, consider a set of $k$ matrices
$\{A_j\}\in\mathbb{C}^{N\times N}$.  Let $\mathcal{S}$ denote the matrix formed by
stacking the $A_j$, so that
\begin{equation}
    S=\begin{bmatrix} A_1\\A_2\\\vdots\\A_k\end{bmatrix}\,.
\end{equation}
Since $\mathcal{S}^\dagger \mathcal{S} = \sum_j A_j^\dagger A_j$, we have that
if $\mathcal{S}\in
V_{N}(\mathbb{C}^{kN})$, then the matrices $A_j$ are valid Kraus operators for a
CPTP map.  Thus we have that there is a correspondence between Kraus operators
and columns of unitary matrices, as noted in \cite{bruzda2009random}.  We will
consider CPTP maps defined by $N^2$ Kraus operators (which is always possible)
and call the associated matrix $\mathcal{S}\in V_{N}(\mathbb{C}^{N^3})$ the
\textit{Stiefel} form of a CPTP map.  Note that since we can rearrange the
index of the Kraus operators, this representation is not unique.

\section{Random Quantum States}
Quantum states are represented by density operators, which are Hermitian trace
one matrices.  This space is compact and can be sampled uniformly, for example
by uniformly sampling the eigendecomposition $\rho=KDK^\dagger$ of the density
operator $\rho$.  The columns of the matrix $K$ are complex and orthonormal,
and are thus unitary matrices which can be sampled uniformly according to the
method described in Section~\ref{sec:stiefel}.  The matrix $D$ is diagonal and
must have $\Tr D =1$, which amounts to uniformly sampling from the simplex,
which can be accomplished (for example), by sampling $N+1$ numbers uniformly in
$[0,1]$, sorting them, and taking their difference as the sample elements on
the diagonal of $D$ \cite{Devroye1986nonuniform}.  With probability one,
density operators generated in this fashion will have rank $N$, but by setting
some diagonal elements of $D$ to zero, we can achieve an arbitrary purity level
(although for pure states it is more efficient to sample uniformly from from
$V_1(\mathbb{C}^N)$.

Suppose, however, we have performed state tomography on a quantum system and
have produced an (average) estimate for a state, $\hat{\rho}$.  Unless
$\hat{\rho}$ is close to $\mathbbm{1}/N$, then a uniform model is unlikely to be
representative.  Certainly, we would hope that state preparation is very close
to the targeted pure state.  Below, we discuss two classes of distribution that
are capable of producing non-uniform and concentrated distributions of random
states.

\subsection{Random Pure States}

First, consider a random \textit{pure} state with (random) density operator
$\rho$.  Since the space of density operators is convex, we have that the
average $\hat{\rho}$ of random pure states is also a valid density operator,
but it is not necessarily pure (and will only be so if the random variable is
constant).  In this case, a natural representation of the pure state $\rho$ is
a unit vector $\varphi$ called a (generalized) Bloch vector
\cite{nielsen2010quantum} which is also the state representation used in an
affine form of a CPTP map \cite{fujiwara1999one}.  The unit vector $\varphi$ is
an element of $V_1(\mathbb{R}^{N^2-1})$ by definition, and the average
Bloch vector $\hat{\varphi}$ is the sufficient statistic for the vector Von
Mises-Fisher distribution on $V_1(\mathbb{R}^{N^2-1})$.  The vector Von
Mises-Fisher distribution is specified by two natural parameters, the
\textit{mean direction} $\mu$ (a unit vector), and the concentration parameter
$\kappa\in(0,\infty)$.  The probability density function of the vector Von
Mises-Fisher distribution is
\begin{equation}
    p_{VF}(X; \mu,\kappa) = c_{VF}(\mu,\kappa)\exp(\kappa\mu^\top X)
\end{equation}
where $c_{VF}$ is a normalization involving $\Gamma$ and Bessel
functions \cite{mardia2009directional}.  Methods for computing the natural
parameters from an average $\hat{\varphi}$ can be found in
\cite{mardia2009directional}.

\subsection{Random Mixed States}
Consider an element $X\in V_k(\mathbb{C}^N)\subset\mathbb{C}^{N\times k}$, and
denote the (orthonormal, by assumption) columns of $X$ by $\vec{x}_i$,
$i=1,\dots,k$, then 
\begin{equation}
    \begin{aligned}
        \Tr(XX^\dagger) &= \sum_{\ell=1}^k\sum_{i=1}^N\sum_{j=1}^N x_{\ell
        i}x_{\ell j}^*\\
        & = \sum_{i,j=1}^k\langle\vec{x}_i,\vec{x}_j\rangle\\
        & = k\,.\\
    \end{aligned}
\end{equation}
Noting $XX^\dagger$ is Hermitian and positive semi-definite (consider
(\ref{eq:choi})), when properly normalized by $1/k$ it is a density density
operator.  Thus, elements $X\in V_k(\mathbb{C}^N)$ can be identified with a
density operator density operators of dimension $N$ with
purity $k$.

Consider the family of probability distributions defined on
$V_k(\mathbb{C}^{N})$ with the following form:
\begin{equation}
    p_{MB}(X;A) = c_{MB}(A)\exp\left(\Tr\left(-X^\dagger A X\right)\right)
\end{equation}
where $c_{MB}(A)$ is a normalizer.  These distributions are known as the
(complex) matrix-Bingham distributions
\cite{mardia2009directional,chikuse2012statistics} and are often used to model
axial data $x$ in which $\pm x$ are indistinguishable, as opposed to
directional data $\theta$, i.e., angles.  These distributions have also found
applications in the analysis of shape \cite{kent1994complex}, where the scale
of the data is normalized and there is a desire for rotation invariance.

Given a Stiefel manifold on $\mathbb{R}^N$ or $\mathbb{C}^N$, the
matrix-Bingham distribution is defined by a natural parameter matrix $A$ whose
dual is the expectation parameter $E[XX^\dagger]$.  This distribution is known
to maximize the Shannon entropy relative to the Haar measure on a given Stiefel
manifold among all distributions with the same moment criterion specified by
$E[X]=0$ and $E[XX^\dagger]$.  Note that the moment constraint $E[X]=0$ is
consistent with the phase ambiguity in quantum mechanics.  Thus, given an
average density operator, we can associate with it a probability distribution
that has sufficient statistics specified by that average, and is in some sense
the least informative distribution on that Stiefel manifold.

Note that the literature is predominately focused on the real-valued matrix
Bingham distribution. However, given $A$ as the natural parameter (i.e.,
concentration matrix) for a complex matrix-Bingham distribution on
$V_k(\mathbb{C}^N)$, let
\begin{equation}\label{eq:complexreal}
	A_R=\begin{bmatrix}\Re A & -\Im A\\\Im A &\Re A\end{bmatrix}
\end{equation}
and let $Y\sim\exp(\Tr Y A_R Y^\top)$ be distributed as a real-valued
matrix-Bingham distribution on $V_k(\mathbb{R}^{2N})$.  Then, let $Y_R$ denote
the first $N$ rows of $Y$ and $Y_I$ denote the remaining $N$ rows.  The random
variable $Z = Y_R+iY_I$ will be distributed as a complex matrix-Bingham
variable with natural parameter $A$ on $V_k(\mathbb{C}^N)$ \cite{kent2013new}.

\section{Random CPTP Maps}
Sampling uniformly from the space of CPTP maps was discussed in
\cite{bruzda2009random} (note their notation is in some sense dual to the
notation here), by first selecting the rank of a Choi matrix and using the
complex Ginibre ensemble with an appropriate normalization. A statistically
identical approach can be achieved by generating $\mathcal{S}$ matrices using
the Ginibre ensemble and the QR decomposition.  In either case, by assuming a
Kraus rank of 1, one can sample uniformly from the space of unitary operations.
Thus, there are known mechanisms for sampling uniformly from various spaces of
CPTP maps.

The focus on fault tolerant quantum computation, however, has emphasized the
generation of gates that are strongly concentrated (in some sense) about the
ideal, so much so that the ensemble average of the random quantum operations
is nearly identical to the ideal gate.  For the purposes of inference (e.g.,
tomography), circuit-level simulation, and system modeling this indicates that we need
distributions on appropriate spaces that are capable of producing strongly
concentrated distributions, and can be defined in terms of the average
CPTP map (i.e., the output of some process tomographic algorithm).

\subsection{Random Unitary Operations and the matrix Fisher
Distribution}\label{sec:cptp_unit}
Let $X$ be an element of $V_N(\mathbb{C}^N)$, as noted above, $X$ is a unitary
matrix, and since representations such as the Choi matrix and Liouvillian
superoperator are inherently quadratic in terms of the entries of $X$, we might
tempted to use the matrix-Bingham distribution with its sufficient statistic
$XX^\dagger$, and look for correspondences to these matrix representations.
However, since $X$ is unitary, $XX^\dagger=\mathbbm{1}_N$ for all $X$, meaning the only
valid matrix-Bingham distribution on this space is equivalent to the uniform
distribution.
 
The next logical step is to consider $V_{N-1}(\mathbb{C}^N)$ to consider random
special unitary operations.  If we decompose $X\in V_{N-1}(\mathbb{C}^N)$ into
its columns by $X=[x_1,x_2,\dots,x_{N-1}]$ and let $\tilde{x}$ denote the
unique vector such that the matrix $[x_1, \dots, x_{N-1}, \tilde{x}]$ is an
element of $SU(N)$.  In this case, we have have that the sufficient statistic
of the matrix-Bingham distribution $XX^\dagger$, can be written as
$\sum_{j=1}^{N-1}x_jx_j^\dagger$.  From the this column-representation, we have
that the expected Choi matrix of the element $SU(N)$ associated with random $X$
is
\begin{equation}\label{eq:su_bing}
    \begin{aligned}
        \Lambda & = E\left[|[x_1,x_2, \dots,x_{N-1},\tilde{x}]\rangle\rangle
        \langle\langle [x_1,x_2,\dots x_{N-1},\tilde{x}]|\right]\\
        &=E\left[\begin{bmatrix}x_1x_1^\dagger & x_1x_2^\dagger &\cdots
            &x_1\tilde{x}^\dagger\\ x_2x_1^\dagger & x_2x_2^\dagger& \dots &
            x_2\tilde{x}^\dagger\\ \vdots &\vdots &\ddots &\vdots\\
        \tilde{x}x_1^\dagger
        &\tilde{x}x_2^\dagger&\cdots&\tilde{x}\tilde{x}^\dagger
        \end{bmatrix}\right]\,.
    \end{aligned}
\end{equation}
From (\ref{eq:su_bing}) we have that the sufficient statistic for a
matrix-Bingham distribution on $V_{N-1}(\mathbb{C}^N)$ can be associated with a
Choi-matrix by summing the first $N-1$ terms on the block-diagonal in
(\ref{eq:su_bing}).  The quantity $E[\tilde{x}\tilde{x}^\dagger]$ is determined
by $\sum_{j=1}^{N-1}x_jx_j^\dagger$ from the partial trace constraints on
$\Lambda$, but the off diagonal terms are essentially being ignored.  In this
sense, the matrix-Bingham distribution is a statistical sub-model of
distributions on $SU(N)$ that have Choi matrices as sufficient statistics, and
information contained in these off-diagonal terms are being discarded.  This is
notionally similar to the relationship between a Guassian random vector with
non-diagonal covariance matrix and its diagonalization.  

In order to capture this missing information in a sufficient statistic, we turn
to the Pauli transfer matrix or affine form of a quantum map.  In this form, a
given quantum operation maps Bloch (or Bloch-like) vectors $\varphi$ to
$A\varphi+\tau$.  In the case of a random unitary map, the channel will be
unital and thus $\tau=0$, meaning the average map is defined only by $A$, and
special unitary operations of the density operator will correspond to special
orthogonal operations on $\varphi$.  Thus, we should look for distributions on
$SO(N^2-1)$ whose sufficient statistics correspond to average elements of
$SO(N^2-1)$. One such candidate is a generalization of the Von Mises-Fisher
distribution called the \textit{matrix-Fisher} distribution.
A random matrix $X$ is said to have the matrix-Fisher distribution with
parameter $F$ if its probability density function (relative to an appropriate
Haar measure) is of the form
\begin{equation}\label{eq:fisher}
    p_{MF}(X;F) = c_{MF}(F)\exp\left(\Tr\left(F^\top X)\right)\right)
\end{equation}
where $c_{MF}(F)$ is a normalizer.

In a similar manner to the way the matrix-Bingham distribution generalizes a
single vector to multiple axes, the matrix-Fisher distribution generalizes
distributions of random angles into random orientations.  Unlike the
matrix-Bingham distribution, the matrix-Fisher distribution distinguishes
between $\pm x$.  The matrix-Fisher distribution is determined by a natural
parameter matrix $F$ whose dual is the expectation parameter $E[X]$, the
additive mean.  Like the matrix-Bingham distribution, the matrix-Fisher
distribution is also a maximum entropy distribution on its associated manifold
with the moment criterion $E[X]$.  

In the literature, there are two subtly distinct families of matrix-Fisher
distributions that generate random elements from $SO(N^2-1)$.  These have the
same general functional form as (\ref{eq:fisher}), but draw from different
sample spaces, namely the Stiefel manifold $V_{n-1}(\mathbb{R}^n)$ and the
rotation group $SO(n)$.  Since $V_{n-1}(\mathbb{R}^n)$ can be uniquely
identified with an element of $SO(n)$, one might think these two distributions
are identical, but they are in fact different.  The parameter space for the
matrix-Fisher distribution on $V_{n-1}(\mathbb{R}^n)$ is a $n\times(n-1)$
matrix, whereas the parameter space for the matrix-Fisher distribution on
$SO(n)$ is a $n\times n$ matrix.  The two families are related, however,
suppose $F_1$ is the natural parameter for the matrix-Fisher distribution on
$V_{n-1}(\mathbb{R}^n)$ and let $F_2=[F_1;\vec{0}]$ be $F_1$ with column of
zeros appended.  The matrix $F_2$ is a valid parameter matrix for the
matrix-Fisher distribution on $SO(n)$, and the two distributions specify the
same distribution of random elements on $SO(n)$.  Thus, the matrix-Fisher
distribution on $V_{n-1}(\mathbb{R}^{n})$ is a statistical sub-model of the
matrix-Fisher distribution on $SO(n)$ (in fact a strict sub-model for $n>2$).
Additional details of the difference between these two classes of distributions
can be found in \cite{sei2013properties}.  Unfortunately, the version of the
matrix-Fisher distribution on $V_{n-1}(\mathbb{R}^{n})$ appears to be the more
well studied version in statistics, but less relevant for quantum applications,
where it seems unlikely that rank-deficient average rotation will be used
often.  That said, using a sign-preserving singular value decomposition, it is
possible to sample from the matrix-Fisher distribution on $SO(n)$ using similar
techniques to the Stiefel manifold version \cite{kent2013new,sei2013properties}.

\subsection{General CPTP Maps}\label{sec:cptp_gen}
As is the case with unitary maps, the matrix-Bingham distribution initially
appears to be an attractive choice for the generation of non-uniform CPTP maps.
Noting that a Choi matrix $\Lambda$ is hermitian, positive semi-definite and
has trace $N$, one might be inclined to use it as the sufficient statistic for a
matrix Bingham distribution, in an identical fashion as the random quantum
state case.  However, even though $\Lambda$ is CPTP, the random output map
$XX^\dagger$ is in general not TP (the use of matrix-Bingham distributions
would be ideal for non-uniform CP maps, however).  One might be tempted to
accept this and hope that the output maps are nearly TP when $\Lambda$
indicates a strongly concentrated matrix-Bingham distribution, and that the TP
normalization process would not affect the distribution too much.   We
attempted to use this technique, and in general it produces CPTP maps that are
not concentrated near the target map.

Another avenue for a statistical model for non-uniform CPTP maps is to note
that the Stiefel representation $\mathcal{S}$ could be generated by a (complex)
matrix von Mises-Fisher distribution.  The complex matrix von-Mises Fisher
distribution can be derived from the real valued distribution via a similar
stacking trick as the Bingham distribution.  This distribution initially shows
more promise than the matrix-Bingham distribution, but it too presents problems
for more subtle reasons.  While a given Choi matrix can be used to generate
Stiefel representation (ignoring for a moment the issue of a non-unique
ordering of the Kraus operators), we have that the average of random Choi
matrix is again a Choi matrix (by convexity), whereas the average of the
equivalent Stiefel representations will lie ``inside'' the Stiefel manifold,
and thus not be the Stiefel representation of a CPTP map. Geometrically, we are
sacrificing convexity for a simple manifold structure.  Thus, using an average
Choi matrix to map to an average Stiefel representation $\mathcal{S}$ will
always result in a degenerate (impulsive) distribution that has all probability
mass at $\mathcal{S}$.  That said, it is conceptually possible to ``scale`` a
given $\mathcal{S}$ by $(1-\varepsilon)$, $\varepsilon\in(0,1)$ that can be
used to define a distribution that generates random Stiefel representations
with concentration about $\mathcal{S}$ parameterized by $\varepsilon$.  
Converting the random Stiefel representations to Choi matrices and taking the
average CPTP map does not empirically appear to be the desired average map, but
it can be made quite close as  $\varepsilon\to0$.  For some use cases this
approximation may be sufficient, and an example application of this approach is
shown in Section~\ref{sec:example:amp}.  It may be possible to compute the
average of the resulting CPTP maps in such a way that the process could be
inverted to find distributions whose average is exactly the desired, but would
still have the issue of Choi matrices being identified with many $\mathcal{S}$.

Having shown that the previously discussed statistical models cannot generate exact
non-uniform samples of arbitrary CPTP maps, we will use the theory of
exponential families to introduce a class of probability distributions
defined on Stiefel manifolds that uses a Choi matrix as its sufficient
statistic.  

Given a set of Kraus operators $A_k$, $k=1,\cdots,N^2$, the entries of the
equivalent Choi matrix $\Lambda$ are 
\begin{equation}
    \Lambda_{ij} = \sum_{k}|A_k\rangle\rangle_i|A_k\rangle\rangle^*_j
\end{equation}
where $|A_k\rangle\rangle_i$ denotes the $i$th entry of the vector
$|A\rangle\rangle$. Next, consider a reshuffling of the rows of the Stiefel
representation $\mathcal{S}$, defined by the same $A_k$ where
\begin{equation}
    \xi = \begin{bmatrix}
        |A_1\rangle\rangle_1&|A_1\rangle\rangle_{N+1}&\hdots&|A_1\rangle\rangle_{N(N-1)+1}\\
        |A_2\rangle\rangle_1&|A_2\rangle\rangle_{N+1}&\hdots&|A_2\rangle\rangle_{N(N-1)+1}\\
        \vdots&\vdots&\ddots&\vdots\\
        |A_N\rangle\rangle_1&|A_N\rangle\rangle_{N+1}&\hdots&|A_N\rangle\rangle_{N(N-1)+2}\\
    |A_1\rangle\rangle_2&|A_1\rangle\rangle_{N+2}&\hdots&|A_1\rangle\rangle_{N(N-1)+2}\\
    \vdots&\vdots&\ddots&\vdots\\
    |A_N\rangle\rangle_N&|A_N\rangle\rangle_{2N}&\hdots&|A_N\rangle\rangle_{N^2}\\\end{bmatrix}\,,
\end{equation}
and decompose this into a block form 
\begin{equation}
    \xi=\begin{bmatrix} \xi_1&\xi_{N+1}&\hdots&\xi_{N(N-1)+1}\\
    \xi_2&\xi_{N+2}&\hdots&\xi_{N(N-1)+2}\\ \vdots &\vdots&\ddots&\vdots\\
    \xi_N&\xi_{2N}&\hdots&\xi_{N^2}\\ \end{bmatrix}\,.
\end{equation}
Then, we have that $\Lambda_{ij}=\xi^\dagger_j\xi_i=\langle
\xi_j,\xi_i\rangle$. Since we only re-shuffled rows, $\xi$ is still an element
of the same Stiefel manifold as $\mathcal{S}$.  Treating an average Choi matrix
$\Lambda$ as a sufficient statistic for $\xi$ is in effect specifying average
inner products between the components in the block structure of $\xi$ in a way
that is not captured by the matrix-Fisher or matrix-Bingham distributions.
Such a distribution would have exponential form 
\begin{equation}\label{eq:schultz_dist}
    p(\xi;\Theta)=C(\Theta)\exp\left(\sum_{i,j=0}^{N-1}
    \begin{bmatrix}\xi_{Ni+1}\\\xi_{Ni+1}\\\vdots\\\xi_{Ni+N}\end{bmatrix}^\dagger
    \mathcal{A}_{i,j}
    \begin{bmatrix}\xi_{Nj+1}\\\xi_{Nj+1}\\\vdots\\\xi_{Nj+N}\end{bmatrix}
    \right)
\end{equation}
where each $\mathcal{A}_{i,j}$ denotes the matrix
\begin{equation}
    \mathcal{A}_{i,j}=\begin{bmatrix}\Theta_{Ni+1,Nj+1}I_N &
        \Theta_{Ni+1,Nj+2}I_N &\cdots &\Theta_{Ni+1, Nj+N}I_N\\
    \Theta_{Ni+2,Nj+1}I_N & \Theta_{Ni+2,Nj+2}I_N &\cdots &\Theta_{Ni+2,
        Nj+N}I_N\\
    \vdots & \vdots&\ddots &\vdots\\
    \Theta_{Ni+N,Nj+1}I_N & \Theta_{Ni+N,Nj+2}I_N &\cdots &\Theta_{Ni+N,
    Nj+N}I_N\\\end{bmatrix}
\end{equation}
and $\Theta$ is the natural form of the parameter (i.e., the dual coordinate
system to the expectation parameter $\Lambda/N$ \cite{amari2007methods}).  For
positive semi-definite $\mathcal{A}_{i,j}$, the distribution in
(\ref{eq:schultz_dist}) is the complex version of a generalized frame-Bingham
distribution \cite{kume2013saddlepoint,arnold2013statistics}.  The real-valued
frame-Bingham distribution can be Gibbs sampled via the techniques of
\cite{hoff2009simulation} as per the discussion in
\cite{kume2013saddlepoint,arnold2013statistics}, and using standard tricks for
converting complex vector operations to real ones (see e.g.,
(\ref{eq:complexreal})) the complex variant can be generated from an
appropriate real-valued distribution.  As far as inference procedures for the
frame-Bingham distribution, \cite{kume2013saddlepoint} introduces a procedure
for approximating the normalizer $C(\Theta)$, but we conjecture that given the
additional structure imposed by $E[\xi\xi^\dagger]=\Lambda\otimes\mathbbm{1}_N$
the estimation process is replicated using the traditional Bingham
distribution.  Showing this explicitly is an area of future research.

\subsection{Other Relevant Distributions}

There are other non-uniform distributions on that appear in the literature that
may be relevant in the context of quantum information.  The matrix-Fisher and
matrix-Bingham distribution are actually subfamilies of the generalized matrix
Bingham-von Mises-Fisher distribution, which has the form
\begin{equation}
    p_{BMF}(X;A,B,C) = c_{BF}(A,B,C) \exp\left(\Tr \left(C^\dagger X+BX^\dagger
    AX\right)\right)\,.
\end{equation}
For spherical data (that is data on $V_1(\mathbb{R}^n)$), there are notions of
bivariate extensions of the Fisher distribution (see \cite[\textsection
11.4]{mardia2009directional} for a brief discussion and references).
Additionally, the frame-Bingham distribution described above is itself a
sub-model of products of the generalized matrix-von Mises-Fisher distribution
\cite{kume2013saddlepoint}.  Such bivariate and multivariate extensions to the
general Stiefel manifolds will likely be needed for analysis of correlated gate
sequences.

\subsubsection{Non-Exponential Families}
All of the families of probability distributions discussed above are
exponential families, with the exception of the uniform distributions (which
are generally limits of exponential families), and as such enjoy the
sufficiency and maximum entropy properties, among a host of other geometrically
motivated properties \cite{amari2007methods}.  There are a number of other
statistical models in the literature that are not exponential families, but
nevertheless have other attractive qualities, and we
will briefly touch on them here.  In general, we are primarily exploiting
isomorphisms in the geometry of a given Stiefel manifold and linking them to
classes of quantum objects.  Thus, any probability distribution defined on a
relevant Stiefel manifold will define random quantum objects of that class, but
may not have as clear of a link in terms of average states or maps.

The matrix angular central Gaussian (MACG) distribution
\cite{chikuse1990matrix} is another model for axial data that is easily
generated from a matrix normal random variate
\cite{gupta1999matrix,chikuse2012statistics}.  Like the matrix Bingham
distribution, it is antipodally symmetric, and is specified by a concentration
matrix $A$ that determines the qualitative shape of distribution on the Stiefel
manifold. Furthermore, MACG and matrix Bingham distributions specified by the
same $A$ will have similar shapes.  In fact, it is used in a rejection sampling
scheme for Bingham random variables \cite{kent2013new}.  In \cite{kent2013new}
the relationship between the matrix Bingham distribution and MACG distribution
is described as analogous to the one between the normal and Cauchy
distributions.  

Another broad class of distributions that are worth mentioning are so-called
wrapped distributions \cite{mardia2009directional}.  For the angular case, this
corresponds to taking some random variable $x$ on the real line, and
considering the angular random variable $\theta=x \text{ mod } 2\pi$.  Common
examples include the wrapped normal and Cauchy distributions.  This concept can
be extended to more general manifolds by considering multi-variate
distributions on the tangent space of the manifold
\cite{chikuse2012statistics}.  Due to the wrapping, it can be difficult to
estimate parameters of the underlying distribution but in the case of the
wrapped normal, the von-Mises distribution and its generalizations has been
shown to be close in shape to the wrapped normal and in fact asymptotically
approach it as the distributions become more concentrated.
Due to particular properties of the Fourier transform of the normal
distribution's probability density function, it satisfies the property that the
sum of two wrapped normals is also a wrapped normal
\cite{jammalamadaka2001topics}, a composition property that the von-Mises
distribution does not satisfy.  Since angular addition is equivalent to the
composition of rotation operations (in this dimension), further exploration of
wrapped distributions in the more exotic spaces considered here would be useful
if we desire a family probability distributions that model random quantum
operations that are closed under composition.

\section{Examples}

\subsection{Dephasing Noise}
Consider a single qubit system with a Hamiltonian of the form
\begin{equation}\label{eq:dephasing}
    H_z(t) = \zeta(t)\sigma_Z
\end{equation}
where $\zeta(t)$ is a stochastic process.  Then, the quantity $U(t)$ defined by
the solution to the differential equation
\begin{equation}
    i\frac{d}{dt}U(t) = H_z(t)U(t)\,,\,\,\,\,U(0)=\mathbbm{1}_N
\end{equation}
is a random unitary operation.  Fix $t=T$ and let $U\triangleq U(T)$.  Since
all of the $H_z$ are proportional to $\sigma_Z$, $U$ will be of the form
\begin{equation}
    U =
    \begin{pmatrix}\alpha&0\\0&\alpha^*\end{pmatrix}
\end{equation}
for some $\alpha\in\mathbb{C}$ with $|\alpha|=1$.  Alternatively, since
$|\alpha|=1$, $\alpha=\exp{i\theta}$, where $\theta =
\int_0^T\zeta(t)\,dt$ for a particular trajectory of $\zeta$.

\begin{figure}[h!]

    \centering

    \begin{tabular}{cc}
        \subfloat[][]{\label{fig:dephasing:a} 
        \includegraphics[width=.4\columnwidth]{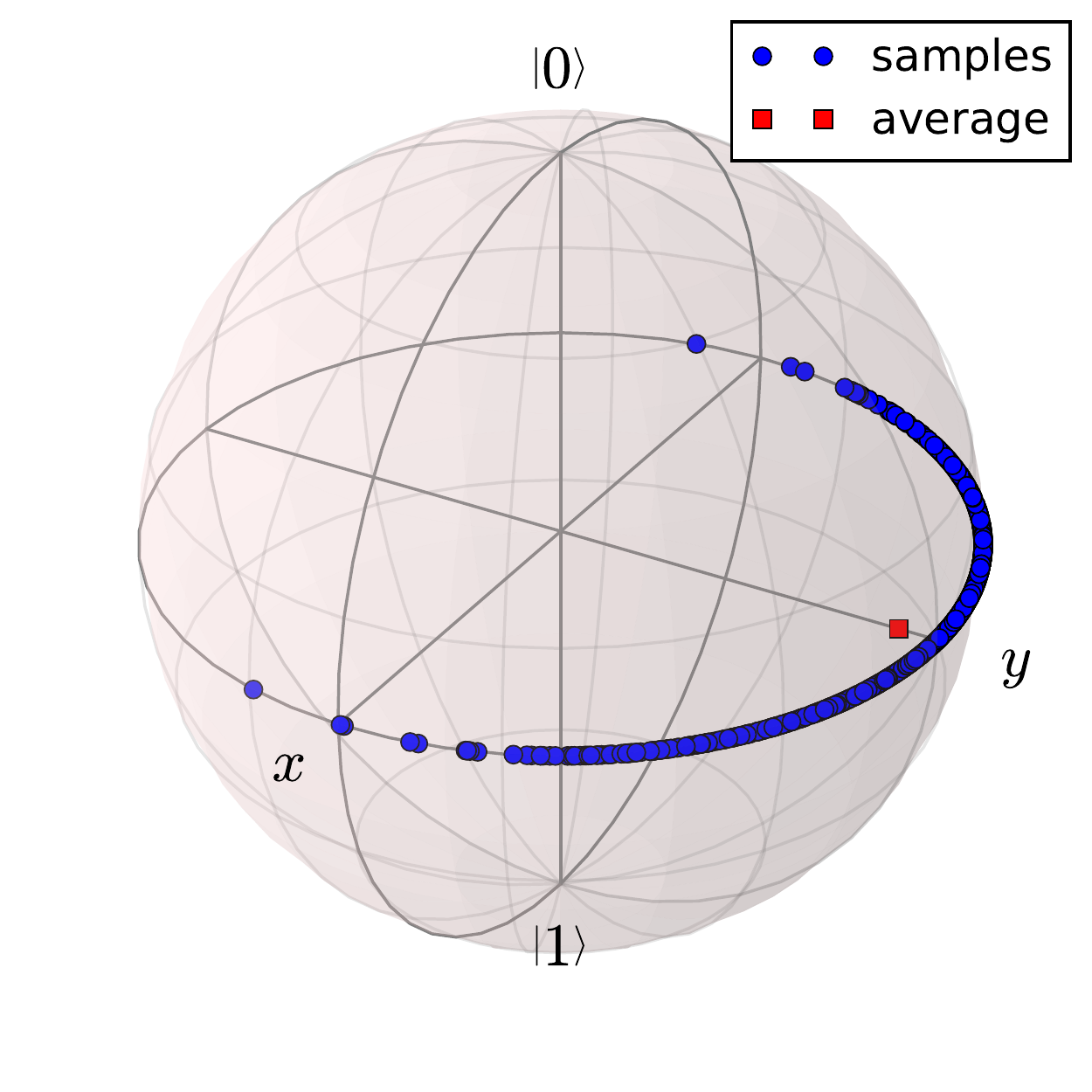}
        }
        &
        \subfloat[][]{\label{fig:dephasing:b} 
        \includegraphics[width=.5\columnwidth]{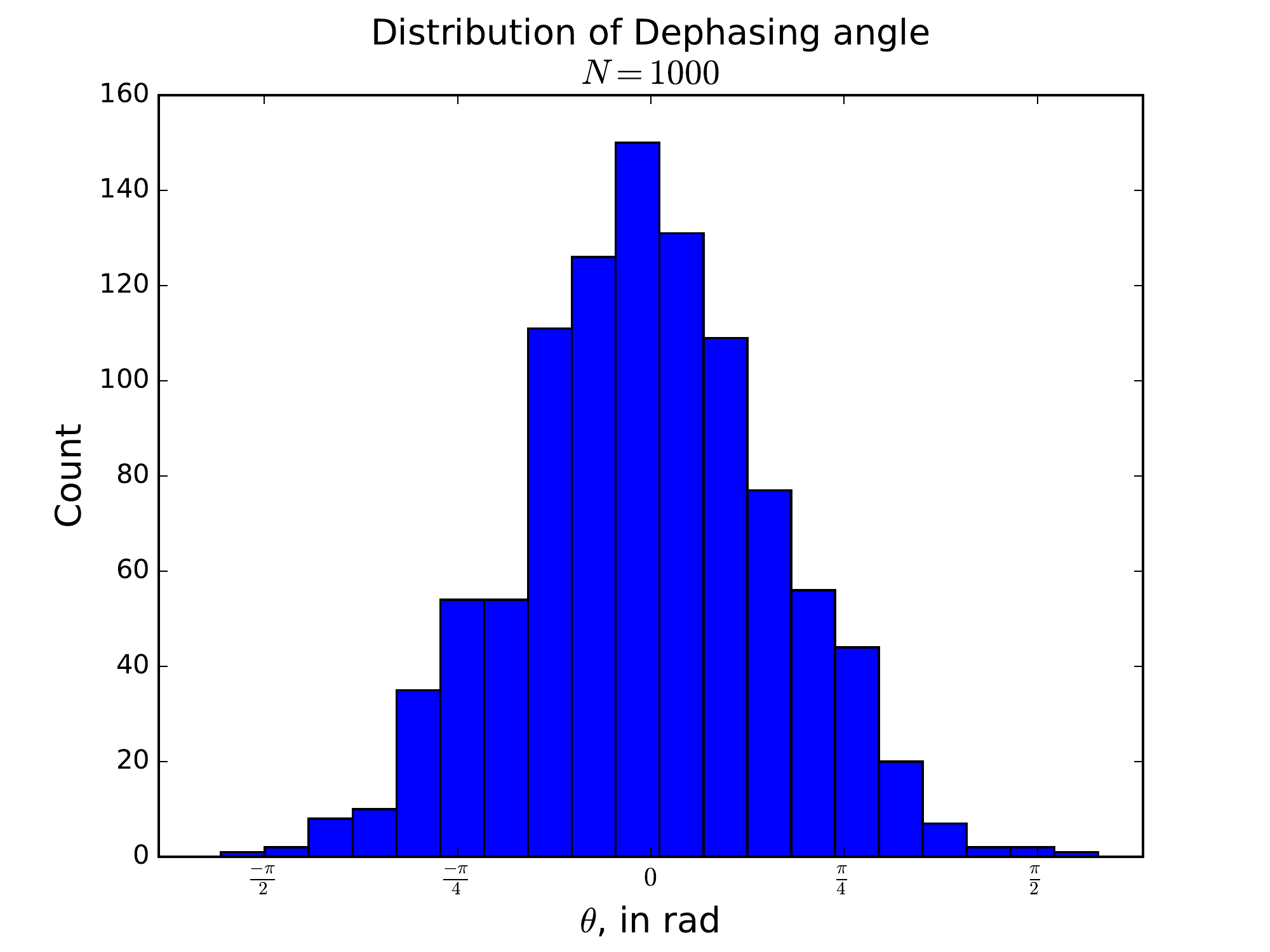}
        }
    \end{tabular}

    \caption{\protect\subref{fig:dephasing:a} Randomly sampled dephasing operations
    drawn from the Von Mises distribution as described above and applied to the
    Bloch vector $[0,1,0]^\top$.  \protect\subref{fig:dephasing:b} Histogram of
    random samples of dephasing angle $\theta$ sampled according to the Von
    Mises distribution corresponding to the data in \protect\subref{fig:dephasing:a}.}

    \label{fig:dephasing}

\end{figure}

The corresponding Liouvillian to $U$ will be of the form
\begin{equation}\label{eq:liou}
    \mathcal{L}=U^*\otimes U = \begin{bmatrix}1&0&0&0\\ 0&\alpha^{*^2}&0&0\\
0&0&\alpha^2&0\\ 0&0&0&1\\\end{bmatrix}\,.
\end{equation}
Let
$T=[|\sigma_I\rangle,|\sigma_X\rangle,|\sigma_Y\rangle,|\sigma_Z\rangle]^\dagger$,
then the Pauli Transfer matrix is
\begin{equation}\label{eq:ptm}
    \mathcal{R} = T^{-1}\mathcal{L}T = \begin{bmatrix}1&0&0&0\\ 0&\Re
\alpha^2&\Im\alpha^2&0\\ 0&-\Im\alpha^2&\Re\alpha^2&0\\ 0&0&0&1\end{bmatrix}\,,  
\end{equation}
which has corresponding affine form
\begin{equation}
    \varphi\to\underbrace{\begin{bmatrix}\Re
    \alpha^2&\Im\alpha^2&0\\ -\Im\alpha^2&\Re\alpha^2&0\\
        0&0&1\end{bmatrix}}_{A}\varphi+\underbrace{\begin{bmatrix}0\\0\\0\end{bmatrix}}_{\tau}
\end{equation}
Let $A$ denote the $3\times3$ sub-matrix in the lower-right corner of
$\mathcal{R}$.  Since $U$ is special unitary, $A\in SO(3)$, and thus $A$ is a random
element of $SO(3)$.  Suppose we are given an \textit{average} dephasing Pauli
transfer matrix $\bar{A}$ generated by the dynamics in (\ref{eq:dephasing}),
then it is 
\begin{equation}
    \bar{A} = \begin{bmatrix}\Re E[\alpha^2]&\Im E[\alpha^2]&0\\ -\Im
E[\alpha^2]&\Re E[\alpha^2]&0\\ 0&0&1\end{bmatrix}\,.  
\end{equation}
To sample random matrices from $SO(3)$ using the matrix-Fisher distribution
with this average $\bar{A}$ is problematic, since it is degenerate, as the
third column must always be $[0,0,1]^\top$.  Instead, the matrix-Fisher
distribution on $SO(2)$ should be used and used to generate the upper-left
$2\times2$ portion of the random $SO(3)$ element.  Furthermore, the
matrix-Fisher distribution on $SO(2)$ is equivalent to the matrix-Fisher
distribution on $V_1(\mathbb{R}^2)$ \cite{sei2013properties} which is in turn
equivalent to the Von Mises distribution on the circle
\cite{mardia2009directional}, although the latter is typically expressed in
terms of an angle $\theta$ rather than an element of $\mathbb{R}^2$.
Figure~\ref{fig:dephasing} shows random dephasing operations generated using
the above scheme, with average dephasing strength $E[\alpha^2] = .9$.

\subsection{Depolarizing noise}
\begin{figure}[h!]

    \centering

    \begin{tabular}{c}
        \includegraphics[width=.4\columnwidth]{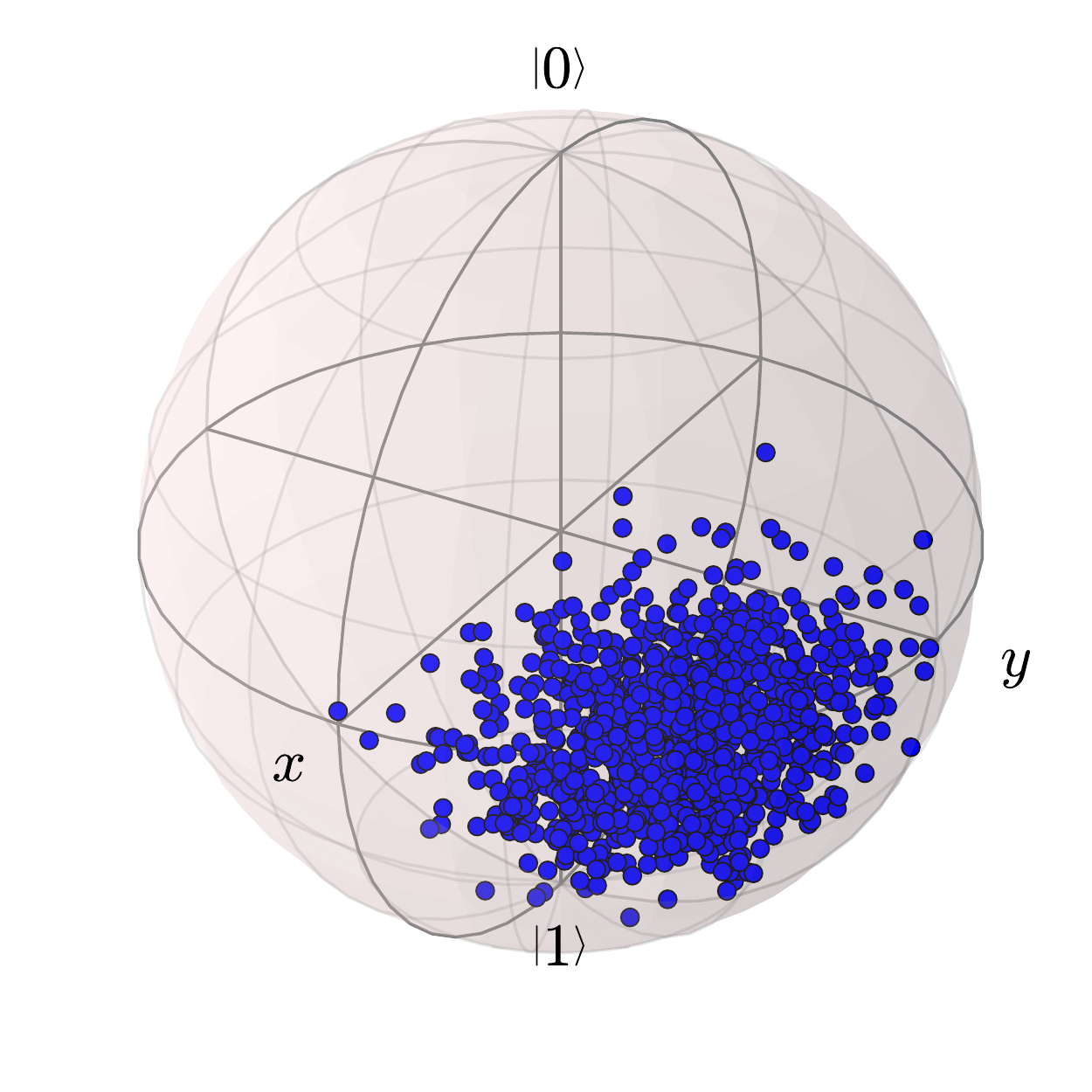}
    \end{tabular}

    \caption{Randomly generated unitary maps whose average is the depolarizing
    channel with $p_x=0.001$, $p_y=0.01$, and $p_z=0.1$ sampled according to the
    Von Mises-Fisher distribution on $SO(3)$ applied to Bloch vector
    $[\frac{\sqrt{2}}{2},\frac{\sqrt{2}}{2},0]^\top$.
    }

    \label{fig:depol1}

\end{figure}

The Choi matrix for Pauli channel depolarizing noise is
$(1-p_x-p_y-p_z)|\sigma_I\rangle\rangle\langle\langle \sigma_I| +
p_x|\sigma_X\rangle\rangle\langle\langle
\sigma_X|+p_y|\sigma_Y\rangle\rangle\langle\langle
\sigma_Y|+p_z|\sigma_Z\rangle\rangle\langle\langle \sigma_Z|$, with
$p_i\in[0,1]$, $\sum p_i\leq 1$.  Converting this to a Pauli transfer matrix
yields
\begin{equation}
    \begin{aligned}
        \mathcal{R} &= (1-p_x-p_y-p_z)I_4\\
        &+p_x\begin{bmatrix}1&0&0&0\\0&1&0&0\\0&0&-1&0\\0&0&0&-1\end{bmatrix}
            +p_y\begin{bmatrix}1&0&0&0\\0&-1&0&0\\0&0&1&0\\0&0&0&-1\end{bmatrix}
                +p_z\begin{bmatrix}1&0&0&0\\0&-1&0&0\\0&0&-1&0\\0&0&0&1\end{bmatrix}
    \end{aligned}
\end{equation}
so the corresponding affine form is
\begin{equation}\label{eq:depol_affine}
    \varphi\to \underbrace{\begin{bmatrix} 1 -2(p_y+p_z)&0&0\\0&1-2(p_x+p_z)&0\\
        0&0&1-2(p_x+p_y)\end{bmatrix}}_{A_p}\varphi+\begin{bmatrix}0\\0\\0\end{bmatrix}\,.
\end{equation}
Note that this is consistent with (\ref{eq:ptm}) under the assumption that
$\zeta(t)$ in (\ref{eq:dephasing}) is zero mean so that $E[\Im\alpha^2]=0$.  If
we treat the contraction matrix $A_p$ in (\ref{eq:depol_affine}) as an average of random
elements in $SO(3)$, we can apply the matrix-Fisher distribution on
$SO(3)$. Figure~\ref{fig:depol1} shows random samples of depolarizing
channels drawn from the matrix-Fisher distribution on $SO(3)$ whose average
corresponds to $p_x=0.001$, $p_y=0.01$ and $p_z=0.1$.   

\begin{figure}[h!]

    \centering

    \begin{tabular}{cc}
        \subfloat[][]{\label{fig:depol2:a} 
        \includegraphics[width=.4\columnwidth]{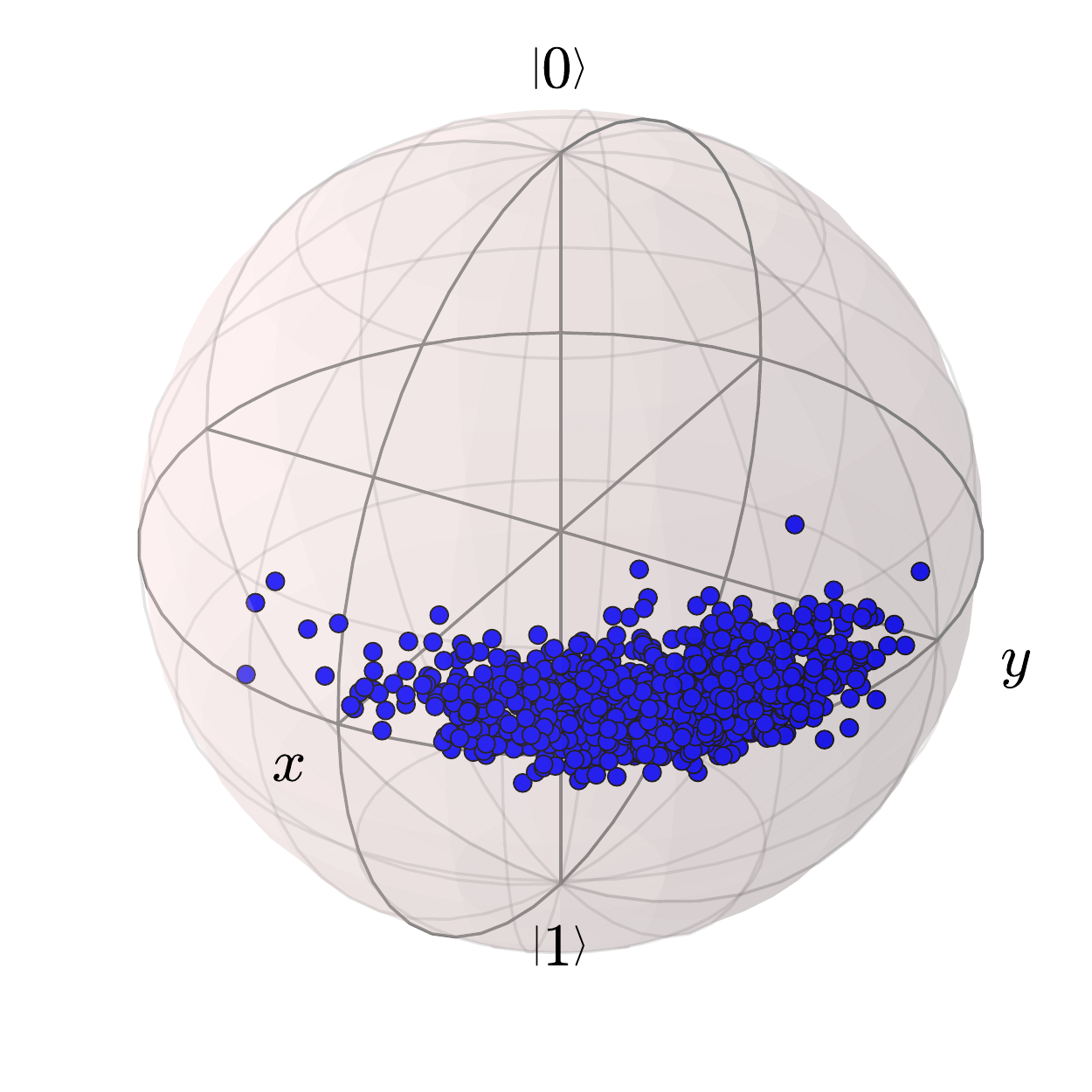}
        }
        &
        \subfloat[][]{\label{fig:depol2:b} 
        \includegraphics[width=.5\columnwidth]{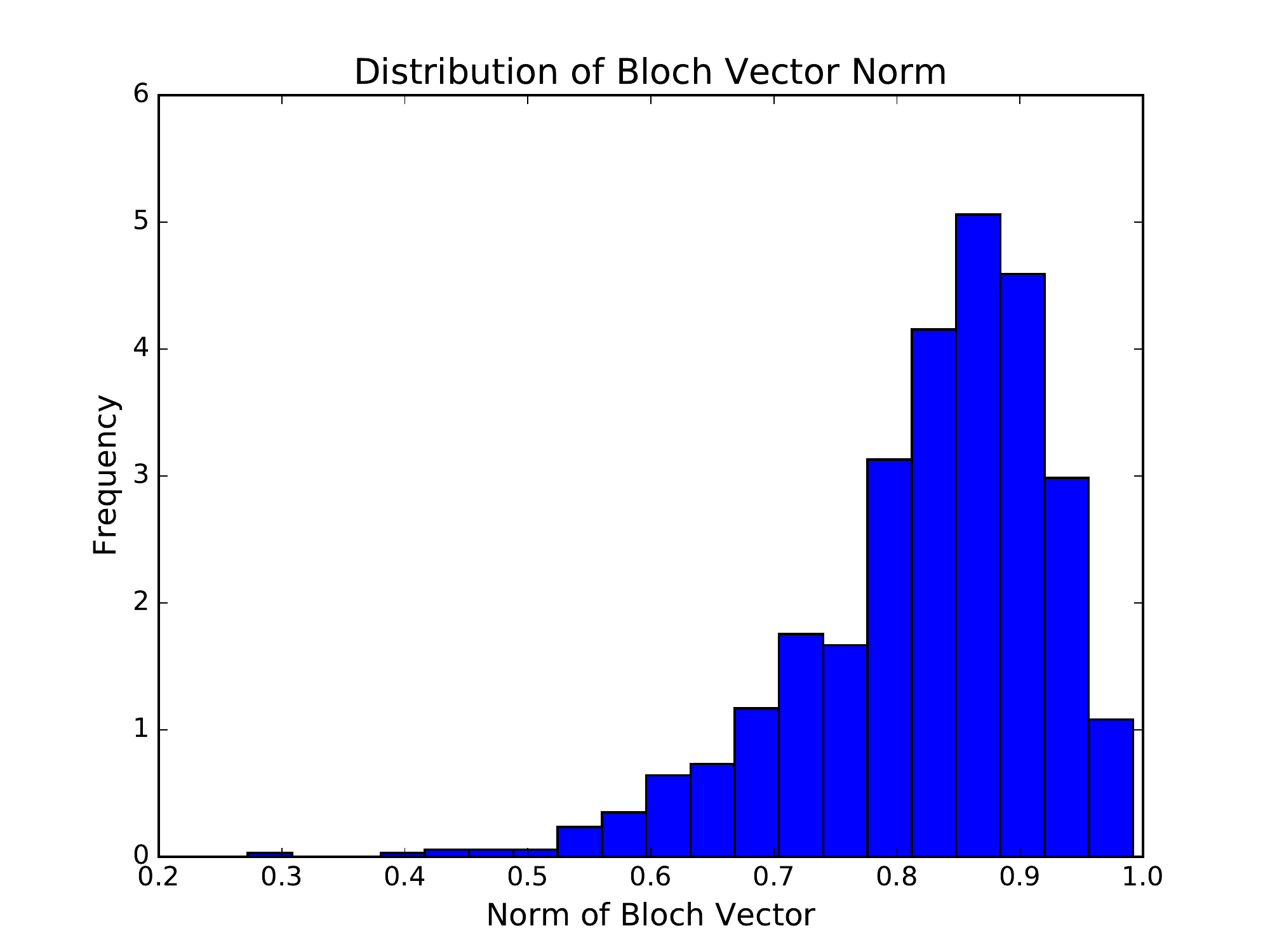}
        }
    \end{tabular}

    \caption{\protect\subref{fig:depol2:a} Randomly generated CPTP maps whose
    average is the depolarizing channel with $p_x=0.001$, $p_y=0.01$, and
    $p_z=0.1$ sampled according to the frame-Bingham distribution applied to
    Bloch vector $[\frac{\sqrt{2}}{2},\frac{\sqrt{2}}{2},0]^\top$.
    \protect\subref{fig:depol2:b} Histogram of norms of output Bloch vectors
    corresponding to the data in \protect\subref{fig:depol2:a}.}

    \label{fig:depol2}

\end{figure}

Since the depolarizing channel is expressed as a mixture over unitary maps, it
is natural to treat a depolarizing channel as the average of random unitary
operations as in Figure~\ref{fig:depol1}.  However, the same depolarizing
channel in Choi form could serve as the sufficient statistic for a
frame-Bingham distribution as described in Section~\ref{sec:cptp_gen}.  In
this case, the random CPTP maps generated would be non-unitary with
probability 1.  Figure~\ref{fig:depol2:a} shows random samples generated
from the frame-Bingham distribution applied to the same input state as
Figure~\ref{fig:depol1}.  Note that the shapes of the output distributions
are are distinctly different.  Furthermore, although it cannot be seen from
the figures themselves, since the operations in Figure~\ref{fig:depol1} are
unitary, the output states are all on the surface of the Bloch sphere.
However, in the case of the frame-Bingham distribution, the states are
pulled in to the Bloch sphere, resulting in a distribution of norms of
Bloch vectors as in Figure~\ref{fig:depol2:b}.

\subsection{Amplitude Damping}{\label{sec:example:amp}
An example amplitude damping channel \cite{nielsen2010quantum}, parameterized
    by $\gamma$ is defined by Kraus operators
\begin{equation}
    \begin{aligned}
        A_1 &= \begin{bmatrix} 1&0\\0&\sqrt{1-\gamma}\end{bmatrix}\,,&&&
        A_2 &= \begin{bmatrix} 0&\sqrt{\gamma}\\0&0\end{bmatrix}\,.
    \end{aligned}
\end{equation}
The corresponding PTM is
\begin{equation}
    \mathcal{R} =
    \begin{bmatrix}1&0&0&0\\0&\sqrt{1-\gamma}&0&0\\0&0&\sqrt{1-\gamma}&0\\
    \gamma&0&0&1-\gamma\end{bmatrix}
\end{equation}
which the first column indicates is non-unital, meaning the techniques in
Section~\ref{sec:cptp_unit} do not apply. Furthermore, amplitude damping
belongs to a class of CPTP maps that cannot be generated as the average of a
random (i.e., non-constant) CPTP map, as discussed in
Section~\ref{sec:representable}.
That said, we can generate distributions whose average is nearly the amplitude
damping channel by using the Stiefel representation
\begin{equation}
    \mathcal{S} = \begin{bmatrix} 1&0\\0&\sqrt{1-\gamma}\\
    0&\sqrt{\gamma}\\0&0\\ 0&0\\ 0&0\\ 0&0\\ 0&0\\ \end{bmatrix}\,,
\end{equation}
and set $(1-\varepsilon)\mathcal{S}$ (for small $\varepsilon$) as the average
value for complex matrix-Fisher distribution as described in
Section~\ref{sec:cptp_gen}. Figure~\ref{fig:ampdamp1} shows the action of
randomly sampled approximate amplitude damping channels ($\gamma=0.01$) on the
initial state $|0\rangle$, for decreasing $\varepsilon$.  For reference, we
find empirically that for $\varepsilon=0.001$ the diamond norm error between
the average approximate amplitude damping channel and the actual is around
$0.05$.

\begin{figure}[h!]

    \centering

    \begin{tabular}{c}
        \includegraphics[width=.4\columnwidth]{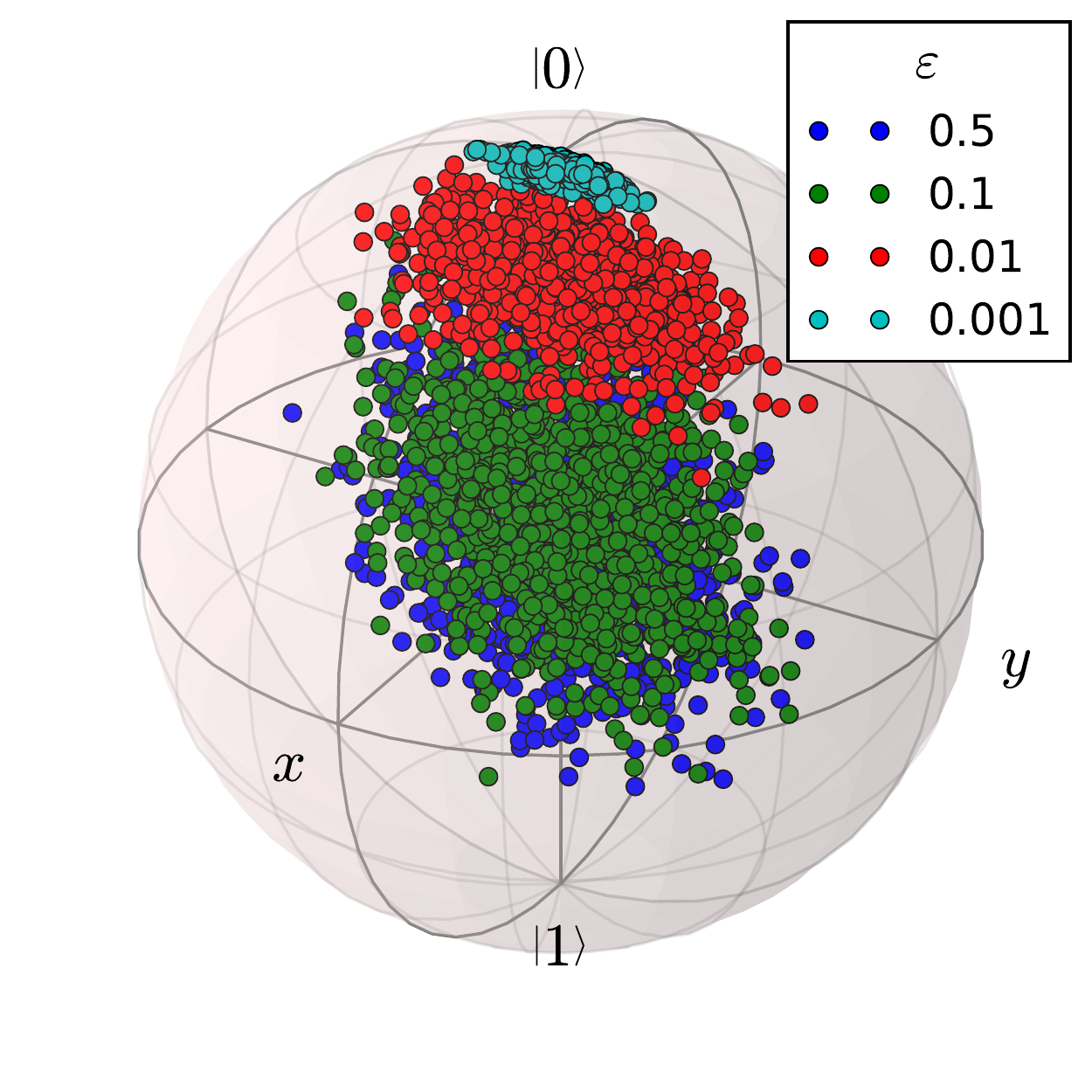}
    \end{tabular}

    \caption{Attempts to approximate an amplitude
    damping channel ($\gamma=0.01)$ using the matrix-Fisher distribution on the Stiefel
    representation for varying $\varepsilon$.  Random samples are applied to
    the state $|0\rangle$ for presentation.}

    \label{fig:ampdamp1}

\end{figure}

In addition to the matrix-Fisher approach, we can also use the frame-Bingham
distribution to attempt to approximate the amplitude damping channel using the
corresponding Choi matrix
\begin{equation}
    \Lambda_\gamma = \begin{bmatrix}1&0&0&\sqrt{1-\gamma}\\
        0&0&0&0\\
        0&0&\gamma&0\\
    \sqrt{1-\gamma}&0&0&1-\gamma\end{bmatrix}\,\,.
\end{equation}
The
first caveat to note is that the Choi matrix for an amplitude damping channel
is in some sense singular, since it indicates that $\xi_2$ in the the
alternative stiefel representation $\xi$ is always 0.  This singularity is
easily handled by a slight modification to the projection step in
\cite{hoff2009simulation} for the first column of $\xi$.  Even with this
additional step, the frame-Bingham distribution is still not capable of
averaging to an amplitude damping channel.  Instead, estimating the parameters
in an attempt match the diagonal terms of the Choi matrix with a frame-Bingham
distribution, while trying to concentrate the distribution as strongly as
possible is shown in Figure~\ref{fig:ampdamp2:a}.  

\begin{figure}[h!]

    \centering

    \begin{tabular}{cc}
        \subfloat[][]{\label{fig:ampdamp2:a} 
        \includegraphics[width=.4\columnwidth]{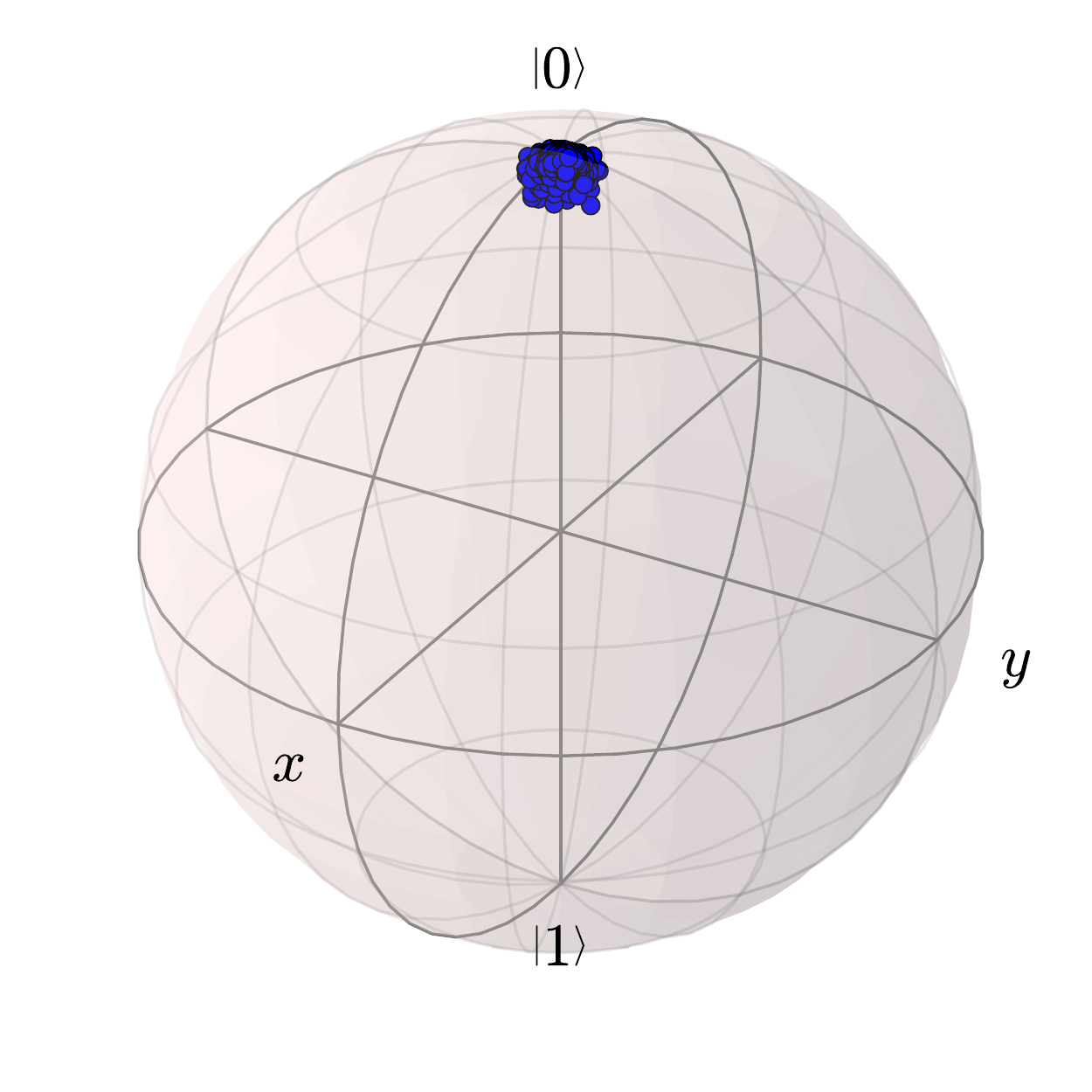}
        }
        &
        \subfloat[][]{\label{fig:ampdamp2:b} 
        \includegraphics[width=.4\columnwidth]{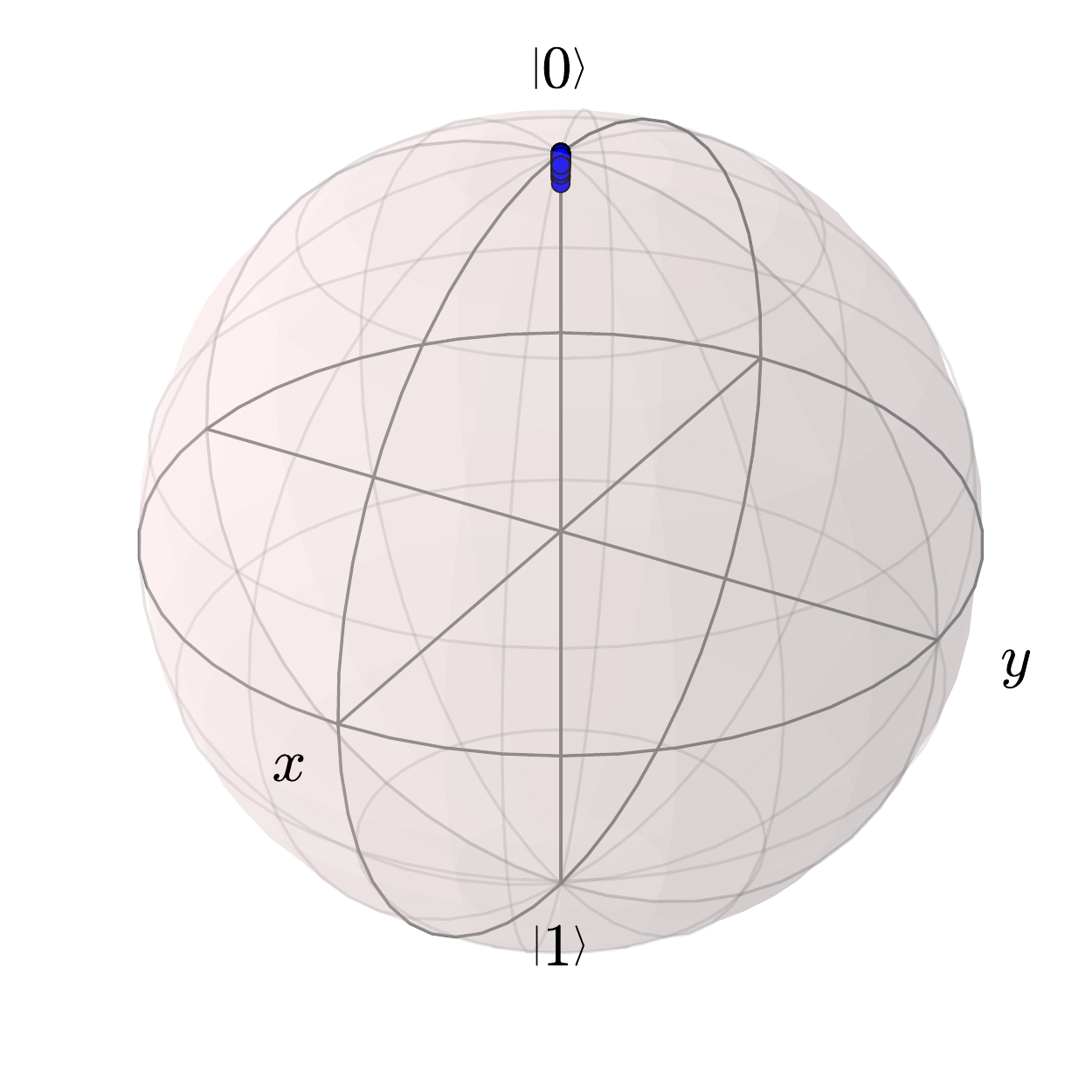}
        }
    \end{tabular}

    \caption{\protect\subref{fig:depol2:a} Randomly generated CPTP maps whose
    average is the depolarizing channel with $p_x=0.001$, $p_y=0.01$, and
    $p_z=0.1$ sampled according to the frame-Bingham distribution applied to
    Bloch vector $[\frac{\sqrt{2}}{2},\frac{\sqrt{2}}{2},0]^\top$.
    \protect\subref{fig:depol2:b} Histogram of norms of output Bloch vectors
    corresponding to the data in \protect\subref{fig:depol2:a}.}

    \label{fig:depol2}

\end{figure}

An alternative approximation that also uses the frame-Bingham distribution is
to further restrict the distribution generated in the projection step to be a
random amplitude damping channel.  Again, the average of random amplitude
damping channels will not be an amplitude damping channel, but we attempt to
approximate by matching the diagonal terms of the Choi matrix.  This approach
is shown in Figure~\ref{fig:ampdamp2:b}.  In this case, since every operation
produced is an amplitude damping channel for some random $\gamma$, the map of
the state $|0\rangle$ lies on the axis, unlike the other approximation cases.

\subsection{A Non-Unital Example}
Consider the amplitude damping channel above followed by the depolarizing
channel.  In PTM form, this channel is
\begin{equation}\label{eq:nonunitalexample}
    \mathcal{R} = \begin{bmatrix}1&0&0&0\\
        0 & (1-2(p_y+p_z))\sqrt(1-\gamma) & 0 & 0\\
        0 & 0 & (1-2(p_x+p_z))\sqrt(1-\gamma) & 0\\
    (1-2(p_x+p_y))\gamma &0 &0 & (1-2(p_x+p_y))(1-\gamma)\\\end{bmatrix}
\end{equation}
which is non-unital, but unlike the pure amplitude damping channel, it can be
represented by a frame-Bingham distribution as shown in Figure~\ref{fig:nonunital}.

\begin{figure}[h!]

    \centering

    \begin{tabular}{cc}
        \subfloat[][]{\label{fig:nonunital:a} 
        \includegraphics[width=.4\columnwidth]{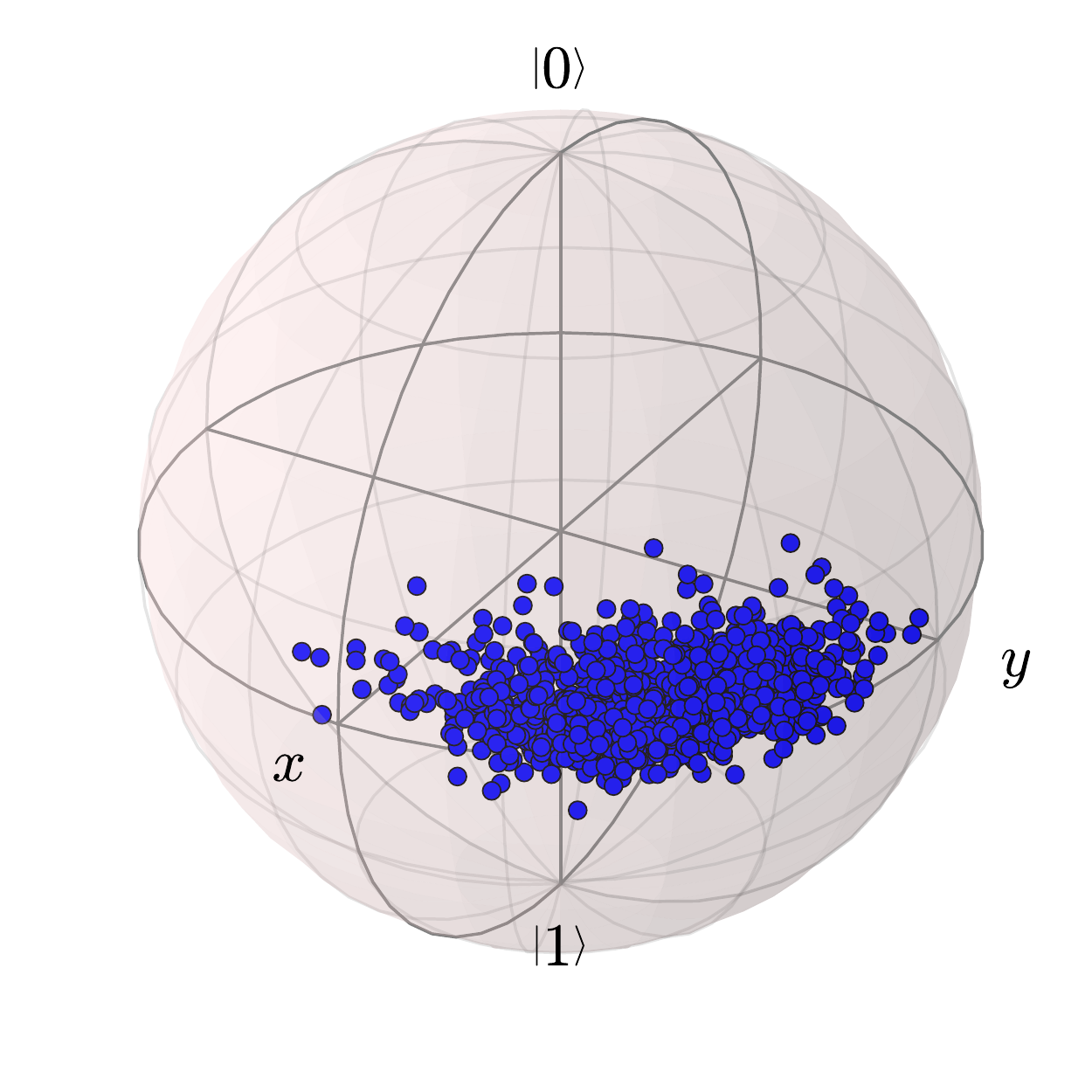}
        }
        &
        \subfloat[][]{\label{fig:nonunital:b} 
        \includegraphics[width=.5\columnwidth]{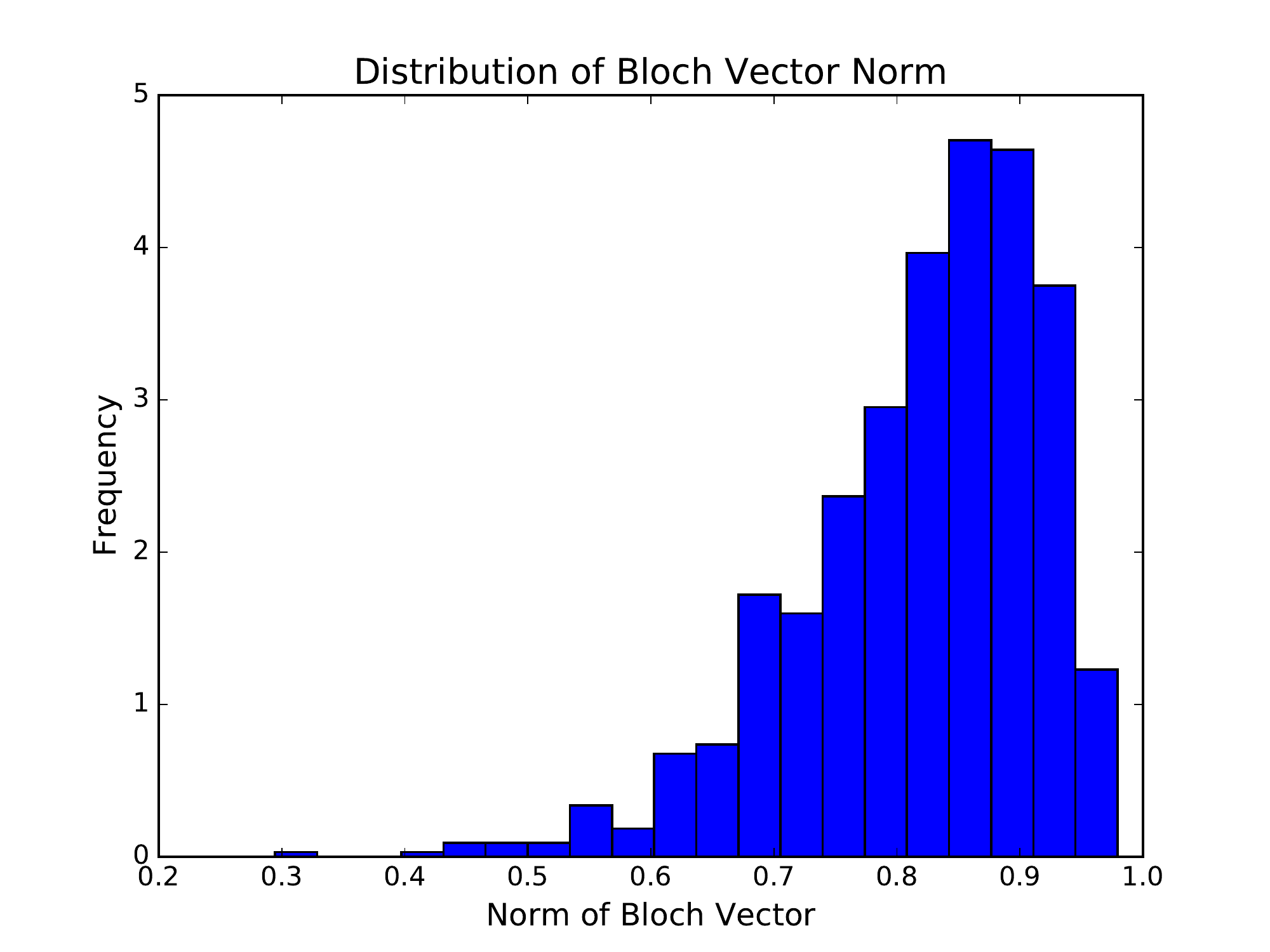}
        }
    \end{tabular}

    \caption{\protect\subref{fig:depol2:a} Randomly generated CPTP maps whose
    average is the nonunital channel in (\ref{eq:nonunitalexample}) with
    $\gamma=0.01$, $p_x=0.001$, $p_y=0.01$, and
    $p_z=0.1$ sampled according to the frame-Bingham distribution applied to
    Bloch vector $[\frac{\sqrt{2}}{2},\frac{\sqrt{2}}{2},0]^\top$.
    \protect\subref{fig:depol2:b} Histogram of norms of output Bloch vectors
    corresponding to the data in \protect\subref{fig:nonunital:a}.}

    \label{fig:nonunital}

\end{figure}

\section{CPTP Maps Representable as Random CPTP Maps}\label{sec:representable}

In this section we will give a geometric characterization the set of CPTP maps
(of a given size) which can be represented by the average of a random CPTP map
of the same size.  Note that this applies to any method of generating random
CPTP maps beyond what is presented here, including Pauli-channel error models
and stochastic master equation approaches.  
For a convex set $\mathcal{C}$, a point $x\in\mathcal{C}$ is said to be an
\textit{extreme} point if $\mathcal{C}\setminus\{x\}$ is still a convex set.  This
definition implies that an extreme point cannot be represented as the convex
combination of other points in $\mathcal{C}$, and as such an extreme point $x$
cannot be the average of a non-trivial random variable taking elements in
$\mathcal{C}$.  Note that there is a distinction between extreme points and
\textit{boundary} points of $\mathcal{C}$, all extreme points are boundary
points, but not all boundary points are extreme (consider the edges of a
square, for example).  Points that are not extreme points must be equal to some
convex combination of extreme points, and thus can be trivially represented as
the average over these points when the extreme points are sampled according to
their weight in the convex combination.  Thus, the question of representability
in this manner is an exercise in classifying the extreme points of the space of
CPTP maps.

For general $N$, this appears to be an open question, however for the case where
$N=2$, the convex geometry of CPTP maps has been completely
characterized in \cite{ruskai2002analysis}.  In particular they show

\begin{theorem}[Theorem 13 in \cite{ruskai2002analysis}] 
    If a CPTP map 
    written in Kraus form $\{A_k\}$ requires exactly two Kraus operators, then
    it is either unital or an extreme point.
\end{theorem}

Since the amplitude damping channel requires two Kraus operators and is
non-unital, it must be an extreme point.  Thus, it cannot serve as the average
for any non-trivial random CPTP map operating on $N\times N$ density
operators, and as such we were unable to exactly reproduce the amplitude
damping channel in Section~\ref{sec:example:amp}.  Furthermore, any random CP
map on $N\times N$ density operators that has an amplitude damping channel as
its average, must take values that are not TP with non-zero probability (c.f.
\cite{burgarth2016can}).

\section{Conclusion}
In this manuscript, we have presented a number of connections between
distributions studied in directional and orientation statistics to various
representations of quantum states and quantum operations.  Furthermore, by
connecting the notion of an average quantum state or operation to a sufficient
statistic of an exponential family we are able to define the unique probability
distribution that maximizes entropy while still satisfying the target average.

From a modeling and simulation perspective, we foresee a number of applications
of this work to the study of quantum systems.  The generation of random quantum
states and operations with a specified average that can be sampled in an
absolutely continuous manner (as compared to the standard Pauli error channel)
can have a drastic effect on simulation results for e.g., error correction and
threshold computations \cite{BarnesPaper}.  Additionally, this sort of
statistical foundation allows for the exploration of correlated quantum errors
which will be present in a non-Markovian environment.

From an inference and analytic perspective, we expect that the statistical
models presented here will be useful for developing statistical tests and
analysis.  For example, tests for correlation from orientation statistics could
be adapted to provide tests and measures for non-Markovianity in quantum
channels.  One can also envision adapting concepts such as graphical models
from applied statistics and machine learning to complex inference problems of
relevance to quantum information, such as tomography.

The work here also touches on some interesting questions on the geometry of
quantum operations, for example, further investigation is warranted of the
exact connection between the vastly different geometries of the Stiefel
representation (a differential geometry on a surface) and the convex geometry
of Choi matrices. Such a viewpoint may be useful in characterizing extreme
points in the convex geometry of quantum operations. Additionally, there may be
interesting connections between the information geometry of the exponential
families presented here and traditional concepts from quantum information.  For
example, does the Kullback-Leibler divergence between elements of the
exponential family relate to any known distance measure between quantum
operations or are there alternative information geometries which maximize
expected diamond norm?

\acknowledgments{This project was supported by the Intelligence Advanced
Research Projects Activity via Department of Interior National Business Center
contract number 2012-12050800010. The U.S. Government is authorized to
reproduce and distribute reprints for Governmental purposes notwithstanding any
copyright annotation thereon. The views and conclusions contained herein are
those of the authors and should not be interpreted as necessarily representing
the official policies or endorsements, either expressed or implied, of IARPA,
DoI/NBC, or the U.S. Government.}

\bibliographystyle{apsrev4-1}
\bibliography{references}

\end{document}